\documentclass[%
 twocolumn, 
 amsmath,amssymb,
 aps,
floatfix,
]{revtex4-2}

\usepackage{graphicx}
\usepackage{dcolumn}
\usepackage{bm}
\usepackage{color, ulem, comment} 

\newcommand{\rno}{$R$NiO$_3$} 
\newcommand{\sno}{SmNiO$_3$} 
\newcommand{\hsno}{H$_{0.5}$-SmNiO$_3$} 
 
\newcommand{\hhsno}{H$_{1.0}$-SmNiO$_3$} 
\newcommand{\oa}{O$^{\rm ap}$}

\definecolor{green}{rgb}{0.0, 0.5, 0.0}

\definecolor{blue}{rgb}{0.0, 0, 0.8}
\begin{document}


%
\title{Hydrogen-Induced Metal-Insulator Transition \\
Accompanied by Inter-Layer Charge Ordering in SmNiO$_3$}
%

\author{Kunihiko Yamauchi}
\author{Ikutaro Hamada}%
\affiliation{%
Department of Precision Engineering, Graduate School of Engineering, Osaka University, Suita, Osaka 565-0871, Japan
}%

\date{\today}

\begin{abstract}
The microscopic mechanism of the hydrogen-induced metal-insulator transition in SmNiO$_3$ is clarified by means of density-functional theory with the Hubbard $U$ correction. 
%
While 100\% of hydrogen doping per Ni atom has been supposed to be responsible for the metal-insulator transition, we found that 50\% of hydrogen doping  results in an outstandingly stable atomic structure showing the insulating property. The stable crystal structure shows the peculiar layered pattern of charge disproportionation of Ni$^{2+}$ and Ni$^{3+}$ valences together with the  strong Jahn-Teller distortion that causes the $e_g$ orbital state splitting and opens the band gap.   

\end{abstract}

                              
\maketitle

{\it Introduction.---}
Transition-metal oxides represented by $AB$O$_3$-type perovskite show intriguing phenomena as functional materials for ferroelectric, magnetic, catalytic and electrochemical applications. Among them, rare-earth nickelates $R$NiO$_3$ ($R$ : rare-earth element)  attract a lot of attention for the peculiar metal-insulator transition (MIT), triggered by the unstable formal oxidation state of Ni$^{3+}$ cations that split into Ni$^{2+}$ and Ni$^{4+}$ in the low-temperature phase.\cite{co1, co2, co3, co4, co5, co6}  
One of possible applications of nickelates related to the MIT is the nano-scale resistive switching device, termed memristor, recently generating huge interest for its use in neuromorphic applications.\cite{neuro, ramanathan1} 
There is also a potential for other device applications such as photonic devices and colossal magnetoresistance devices.\cite{catalano.review} 

%
Recently, it has been shown that the chemical doping can be used to drive the phase change in the SmNiO$_3$ thin film.\cite{ramanathan1, ramanathan2}
There, it has been demonstrated that the electronic conductivity is decreased by more than eight orders of magnitude 
by doping hydrogen.
It has been proposed that additional electrons donated by H atoms change the ionic valence of Ni from Ni$^{3+}$ to Ni$^{2+}$ and cause the MIT. 
Similarly, Li and Mg ion doping was found to lead to the MIT.\cite{ramanathan3} 
%
From the theoretical side, Yoo and Liao have recently introduced H and Li atoms into SmNiO$_3$ in density functional theory  (DFT) simulation with the Hubbard $U$ correction and have observed a metallic state at low H concentrations and 
the MIT at a high H concentration (Ni:H ratio of 1) with the band gap of 3.0 eV in the ferromagnetic configuration.\cite{yoo1} 
The trend of the chemical-doping-induced MIT in $R$NiO$_3$ series has been further investigated.\cite{yoo2, cuidft2021}
A similar DFT simulation has been performed by Kotiuga and Rabe; however, their result shows the insulating state at all the concentrations of 0, $\frac{1}{4}$, $\frac{1}{2}$, $\frac{3}{4}$, and 1 electron/Ni 
probably due to the Hubbard $U$ correction ($U-J = 4$ eV) in $G$-type antiferromagnetic ordering they considered.\cite{rabe.prm2019} 
While the crystal/electronic structure at 100\% H doping ({\it i.e.}, Ni:H ratio of 1) has been investigated, the low-H-doping region has not been explored much. 
%
Direct quantification of hydrogen in tensile-strained \sno\ thin film has been performed by using nuclear reaction analysis and a significant modification in resistivity has been observed despite a low H concentration.\cite{chen.natcom2019}  While it was explained by an intermediate meta-stable state in the transient diffusion process of hydrogen, it may imply that the MIT occurs with a low H concentration accompanied by the structural distortion caused by H doping.  

To resolve the issue and understand the mechanism of hydrogen-induced MIT, detailed information on the crystal structure and occupation sites of
doped H atoms 
is necessary. 
In particular, it is vital to understand how the crystal structure and the associated electronic structure evolve as the hydrogen concentration increases. 
In this study, we performed an exhaustive simulation for H doped \sno\
by 
considering all the H atom arrangements in a supercell. 

{\it Structural and computational details---}
\begin{figure}[htb]
\begin{center}
\includegraphics[width=7.8cm]{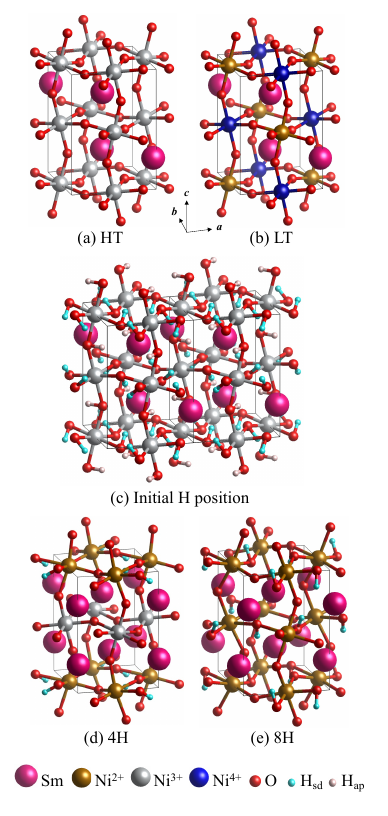}
\caption{  \label{fig:crys} 
 (a) HT $Pbnm$ and  (b) LT $P2_1n$ crystal structures of \sno. (c) Initial H atom positions used in all the H-doping simulations in a 2$\times$1$\times$1 supercell of $Pbnm$ \sno.
 16 side H (H$_\mathrm{sd}$) and 8 apical H (H$_\mathrm{ap}$) sites are shown in different colors.
According to the doping concentration, some of the H sites were chosen to be occupied and other sites were left unoccupied in generating initial H$_x$-\sno\ crystal structures. 
 (d) 4H $Pb$ structure of \hsno\ and (e) 8H $P2_{1}2_{1}2_{1}$ structure of \hhsno\ predicted to be the most stable. 
 }
\end{center}
\end{figure} 
The high-temperature (HT) phase of \rno\ shows the orthorhombic $Pbnm$ crystal structure (non-standard setting of $Pnma$) in which the $c$ axis is the longest one and the $ab$ plane lies parallel to NiO layer [Fig.~\ref{fig:crys} (a)]. With respect to the ideal cubic perovskite,  GdFeO$_3$-type octahedral tilting and the cooperative Jahn--Teller (JT) distortion lowers the symmetry. 
The low-temperature (LT) phase shows monoclinic $P2_1n$ structure as shown in Fig.~\ref{fig:crys} (b). 
In this structure, JT distortion is suppressed while 
the Ni-O bond disproportionation and the breathing-mode distortion of NiO$_6$ octahedra  create two nonequivalent Ni sites (Ni$^{2+}$ and Ni$^{4+}$) in the insulating phase. 
While the LT structure exhibits lower total energy by 
13 
meV than that in HT structure, 
we hereinafter focus on the meta-stable HT structures, as room-temperature \sno\ exhibits a non-disproportionated crystal structure\cite{rtstr1,rtstr2} 
Starting with the HT structure, we now consider the structural distortion caused by H doping into \sno. 
As reported in a previous study\cite{yoo1}, the most preferential site for a single H atom is an interstitial site in the (001) planes near the edges of NiO$_6$ octahedron with O-H bond to the side oxygen (we will call it `side doping' hereafter). Another stable site is found in the (001) plane with O-H bond to the apical oxygen (`apical doping') as shown in Fig.~\ref{fig:crys} (c). 
To systematically investigate the spatial distribution of doped H atoms, we constructed a 2$\times$1$\times$1 $Pbnm$ supercell with four formula unit (f.u.) containing 16 side O and 8 apical O atoms [Fig.~\ref{fig:crys} (c)] and all the possible H patterns were considered by a combinatorial structure-generation approach using the \textsc{Supercell} \cite{supercell} program by increasing the number of the doped H atoms from one to eight.  
This yields the number of combinations explosively increasing as 
$_{16}\rm C_{1}$=16, $_{16}\rm C_{2}$=120, 
$_{16}\rm C_{3}$=560, 
$_{16}\rm C_{4}$=1820, 
$_{16}\rm C_{5}$=4368, 
$_{16}\rm C_{6}$=8008, 
$_{16}\rm C_{7}$=11440, 
$_{16}\rm C_{8}$=12870 for the side doping.
However, these numbers can be reduced to 1, 16, 35, 134, 272, 539, 714, 854, respectively, by taking into account the 16 symmetry operations in the $Pbnm$ space group. In total, 2565  and 42 structures were prepared for the side doping and the apical doping, respectively. 
%
To evaluate the structural stability, 
we used the projector augmented wave method \cite{PAW} as implemented in the \textsc{VASP} code\cite{VASP}.
The Perdew-Burke-Ernzherof (PBE) \cite{PBE} generalized gradient approximation was used for the exchange-correlation and the Hubbard-$U$ correction \cite{dudarev} (PBE+$U$) with the effective $U$ of 3 eV was applied to the Ni 3$d$ state.
Atomic positions and lattice parameters were fully optimized for each H atomic configuration.
%
After the full geometry optimization of the atomic positions and the lattice parameters, 
 the total energy was evaluated by using the tetrahedron method with Bl\"ochl corrections 
\cite{blochl.andersen} 
with $2\times 4 \times 3$ $\bm k$-mesh points. 
During the structural optimization, H atoms freely move around but keeping the bonding to O atoms.  
The spin configuration was set in the ferromagnetic state for simplicity. 
Maximally localized Wannier functions (MLWFs) were calculated by using the WANNIER90 code \cite{wannier1, wannier2, wannier90v2} interfaced with the \textsc{VASP} code.
To characterize the H-induced structural distortion in \sno, the crystalline symmetry and structural distortion modes were investigated by using \textsc{ISOCIF} and \textsc{ISODISTORT} programs.\cite{isocif, isodistort}  
The crystallographic figures were generated using the \textsc{VESTA} program\cite{vesta}. 

{\it Structural stability under hydrogen doping---}
\begin{figure}[htb]
\begin{center}
\includegraphics[width=8.5cm]{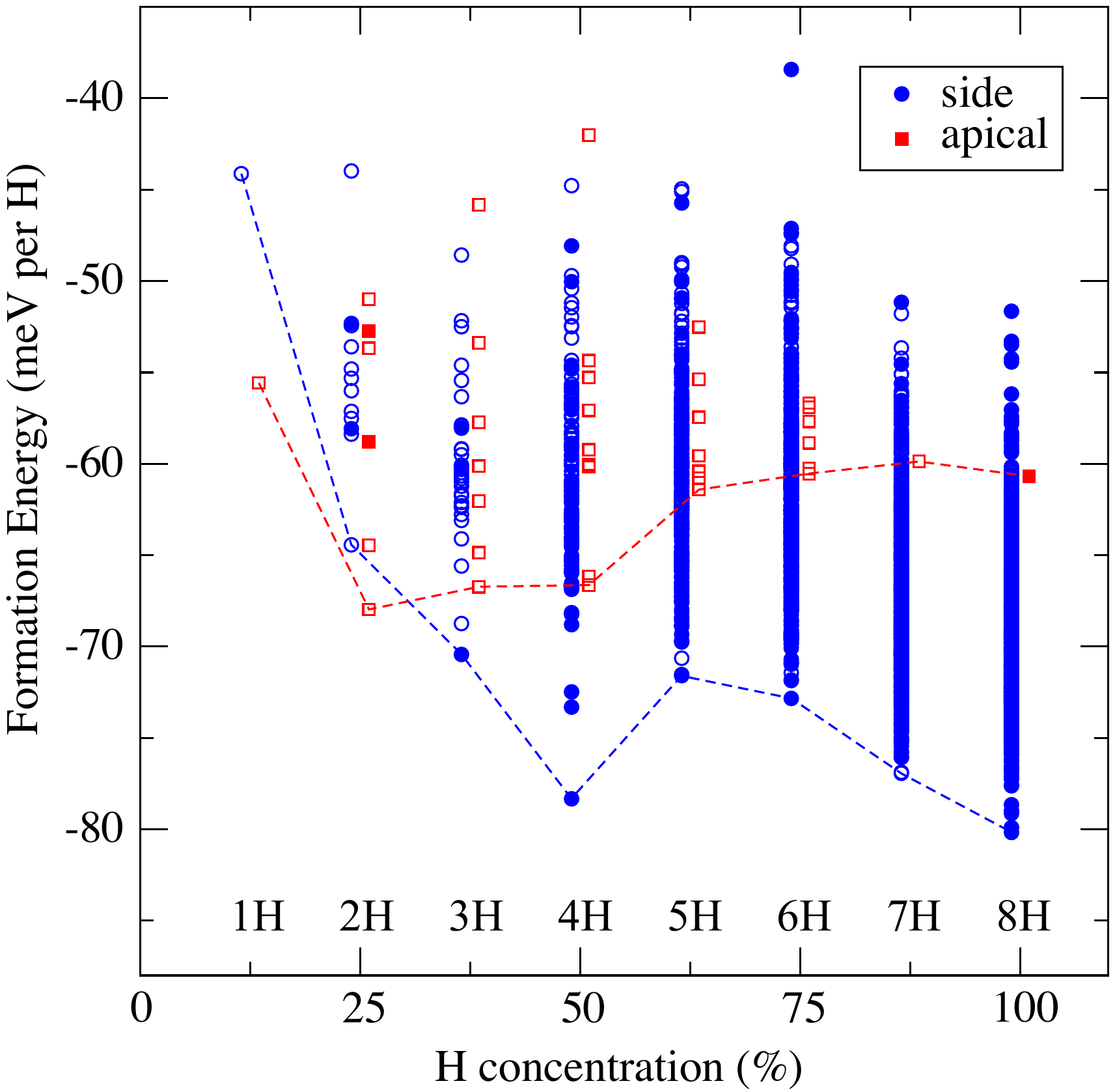}
\caption{\label{fig:en} 
Formation energy of H$_x$-\sno\ as a function of H doping concentration per Ni for side doping and apical doping. Open and closed symbols denote metallic and insulator states, respectively. 
}
\end{center}
\end{figure} 
Figure~\ref{fig:en} shows the calculated formation energy defined as $E_{\rm f}$ = $E$(H$_x$-SmNiO$_3$) $-$ $E$(SmNiO$_3$) $-$  $\frac{x}{2} E$(H$_2$), as a function of H doping concentration $x$. 
The three terms on the right hand side of the above equation denote the total energies of the hydrogen-doped \sno, pristine \sno, and H$_2$ molecule, respectively. 
It is 
found 
that the apical doping shows lower energy at low doping ($\leq$25\%), while the side doping shows lower energy at high doping ($\geq$27.5\%). 
The global energy minimum is located at 100\% (H:Ni=1:1) doping at which the most stable structure shows the formation energy of 
$-$80 
meV. The ratio of insulating structures (those with a finite band gap) increases as the H concentration increases. At 100\% H-doping, all 854 structures show the insulating state. 
Unexpectedly, a prominent local minimum appears at 50\% H-doping 
at which the most stable structure shows the formation energy of
$-$78 meV, more than 5 meV 
lower than the second most stable structure. 
Hereinafter, 
the stable structures are referred to as  ``8H'' and ``4H'', respectively, and we will focus on their structural and electronic properties since these structures may be responsible for the experimentally observed MIT. 

Our symmetry analysis revealed that the 8H structure shows the chiral $P2_12_12_1$ space group and the primitive cell is reduced to a half cell (4 f.u.) as shown in Fig.~\ref{fig:crys}(e). 
Remarkably, the structure has only one equivalent site for Ni and Sm atoms, and three for O atoms. 
Each NiO$_6$ octahedron makes bonding with two H atoms that provide one electron to one Ni atom, being consistent with what was reported in a previous DFT study.\cite{yoo1} 
By comparing the present structure and that in the literature, we found that the newly found structure has higher symmetry (the previous one has $P1$ symmetry) and shows much lower energy by 
5 meV/f.u. 
after optimizing both structures. This finding of the high symmetric structure demonstrates the superiority of the computational approach taking into account all the doped H atom configurations exhaustively.
The band gap is calculated to be 1.41 eV  with GGA+$U$
.  
By using the Heyd-Scuseria-Ernzerhof (HSE06) screened hybrid functional method\cite{hse06}, the calculated band gap increases to be 5.12 eV. 

\begin{figure}[htb]
\begin{center}
\includegraphics[width=8cm]{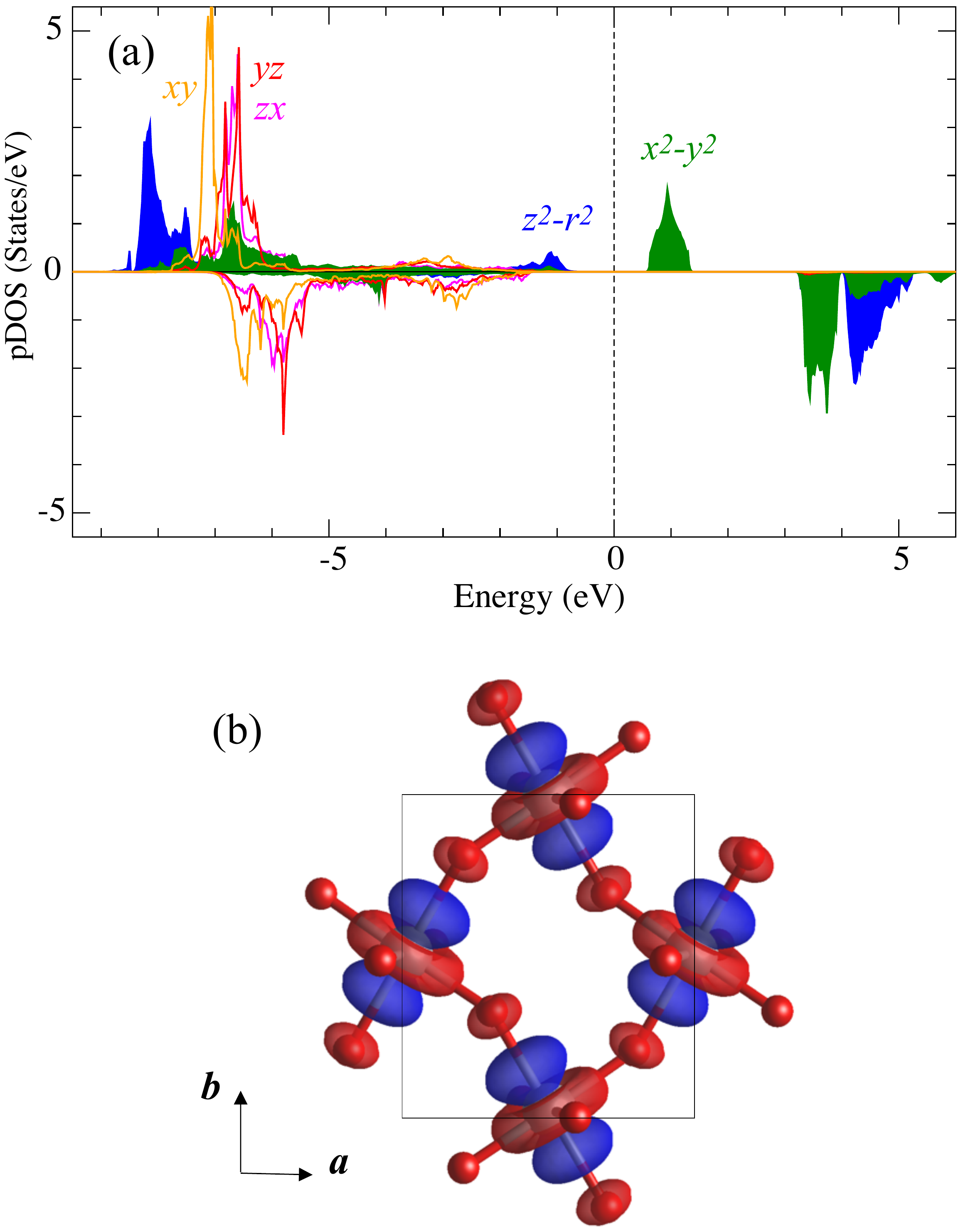}
\caption{\label{fig:wdos} 
(a) Projected DOS of Ni 3$d$ orbital state at Ni$^{3+}$ site 
in 4H structure of \hsno\ using HSE06 exchange-correlation functional.
(b) MLWF contour plot of occupied Ni $e_g$ orbital state 
in the Ni$^{3+}$O$_2$ layer. 
}
\end{center}
\end{figure} 

To our surprise, 
it is found that the 4H structure of \hsno\ also shows the highly symmetric structure in 4 f.u. cell with the $Pb$ symmetry as shown in Fig.~\ref{fig:crys}(d). 
There are two equivalent Ni sites, namely Ni$^{2+}$ and  Ni$^{3+}$, showing the layered charge-ordering pattern forming Ni$^{2+}$O$_2$ layer and Ni$^{3+}$O$_2$ layer alternatively. These ionic valences can be distinguished at a glance by the calculated spin momenta; $S(\rm{Ni}^{2+})= 1.5~\mu_{\rm B}$ and $S(\rm{Ni}^{3+})= 1.0~\mu_{\rm B}$. 
In the Ni$^{2+}$O$_2$ layer, one additional electron is provided by 
an H atom
to Ni site, making the $t_{2g}$ 
state fully occupied. 
In the Ni$^{3+}$O$_2$ layer, Ni ions stay in a trivalent state, while the NiO$_6$ octahedral tilting and Jahn--Teller (JT) distortion are enhanced by H doping. 
This results in the large energy splitting of $e_g$
state as shown in Fig.~\ref{fig:wdos} (a); here the density of states (DOS) is projected onto Ni-3$d$ orbital state under $O_h$ crystal field using MLWFs. 
The band gap is evaluated as 0.36 eV and 1.4 eV by PBE+$U$ and HSE06 methods, respectively. 
In Fig.~\ref{fig:wdos} (b), the orbital ordering can be seen in Ni$^{3+}$O$_2$ layer; the 
alternating $x^2-r^2/y^2-r^2$ orbital ordering pattern looks similar to that caused by cooperative JT distortion in $R$MnO$_3$.\cite{kimura, yamauchi.homno3}  
Making a keen contrast to other perovskite transition-metal oxides, 
the unusual inter-layer charge ordering and the intra-layer orbital ordering coexist in the 4H structure. 

\begin{table}[htb]
    \centering
    \begin{tabular}{|c|c|c|c|c|l|}
    \hline
        & HT    & LT    & 4H    & 8H    & \\
        \hline
    $a$ & 5.262 & 5.279 & 5.230 & 5.386&lattice constants\\
     $b$ & 5.530 & 5.429 & 5.737 & 6.010& (\AA)\\
     $c$ & 7.491 & 7.518 & 7.553 & 7.512&\\
     \hline
     $\alpha$ &90.00  & 90.00  & 90.53 & 90.00 & monoclinic angle ($^\circ$)\\
    \hline
    $V$     & 214.72 & 215.11 & 226.68 & 243.25& cell volume (\AA$^3$/2f.u.)\\
    \hline
    \hline
               & HT    & LT    & 4H    & 8H    & \\
     \hline
    $X_1^+$ &       &       &   0.21&       &interlayer breathing   \\
    $X_5^+$ &   0.74&   0.65&   0.93&   1.13&$a^+$ tilting   \\
    $X_5^-$ &       &       &   0.14&   0.26&bending*\\
    \hline
    $M_2^+$ &  0.15 &   0.01&   0.08&   0.07&JT\\
    $M_2^-$ &       &       &       &   0.25&Sm $c$ displacement\\
    $M_3^+$ &  1.01 &   0.96&   1.36&   1.29&$c^{+}$ tilting \\
    $M_3^-$ &       &       &       &   0.16&\oa\ $c$ displacement\\
    $M_5^-$ &       &       &   0.19&       &\oa, Ni  off-centering\\
    \hline
    $R_1^+$ &       &   0.13&       &       &breathing* \\
    $R_3^+$ &       &   0.00&   0.23&       &JT* \\
    $R_4^+$ &1.35   &   1.31&   1.42&   1.93&$b^+$ tilting* \\
    $R_4^-$ &       &        &  0.20&   0.39&Ni off-centering* \\
    $R_5^+$ & 0.17  &   0.14 &  0.22&   0.42&bending* \\
    \hline
    \end{tabular}
    \caption{The optimized lattice parameters, cell volume, and displacive phonon modes and their amplitudes (in \AA) of HT ($Pbnm$), LT ($P2_1/c$) structure of \sno, 4H ($Pb$) structure of \hsno, and 8H ($P2_12_12_1$) structure of \hhsno\ with respect to the ideal cubic perovskite structure ($Pm\bar{3}m$). Displacive modes 
    with their amplitudes larger than 0.1 \AA\ are listed. $\Gamma$ mode is not listed for simplicity. The asterisk symbol ($\ast$) denotes the out-of-phase mode, while the others are the in-phase mode. }
    \label{tbl:distortion}
\end{table}

Table~\ref{tbl:distortion} shows the amplitude of displacive phonon modes of HT, LT, 4H, and 8H structures with respect to the cubic perovskite structure. The ionic displacement for each mode is illustrated in Supplementary Material. In the HT structure the octahedron-tilting $X_5^+$, $M_3^+$, and $R_4^+$ modes and the $M_2^+$ mode related to JT distortion are dominant. The H atom that makes a bond to the side O atom enhances the octahedron-tilting modes and bending modes in 4H and 8H structures. While the $M_2^+$ mode is suppressed, the $R_3^+$ mode shows a large amplitude in the 4H structure, leading to the out-of-phase JT distortion. These unusual distortion modes are supported by the lattice expansion due to interstitial H doping. 
In the 4H structure, since H atoms are located to form stripes along the $b$ axis, the lattice is expanded along the $b$ direction. This uniaxial strain in turn supports the $R_3^+$ mode softening. 
To highlight the effect, we repeated H-doping simulations in \sno\ in which lattice parameters were fixed as those of a non-doped structure. 
This resulted in a monotonous behavior of the formation energy with respect to the H concentration and the energy minima at 4H and 8H structures disappeared (see Supplemental Material);  
the $R_3^+$ distortion in the 4H structure decreased the amplitude from 0.23 to 0.07 \AA\ in the fixed lattice calculation. 
Therefore, the H doping in nickelate exhibits two important ingredients for the MIT: low-symmetry ionic distortions and lattice expansion that in turn promotes the JT distortion in Ni$^{3+}$O$_2$ layer.

{\it Summary---}
By means of combinatorial DFT calculations, we examined the total energy of more than 2000 structures in  H-doped \sno\ and found a significantly stable structure at the 50\%\ 
H 
concentration manifesting  an unusual layered pattern of charge ordering in the insulating state. 
While our simulation results may provide a reasonable explanation for the experimentally observed MIT,  further experimental studies and structural analysis by quantification of the H doping concentration are needed to confirm 
our findings. 
This
study also demonstrates that a combinatorial structural generation can be a fine tool to find stable structures under chemical doping, 
thus providing a reference for future research in the field of hydrogen-doped oxides, where a significant change in the electronic structure caused by structural distortion could play a fundamental role in tuning the system functionality.


This work was supported by Grant in Aid for Scientific Research on Innovative Areas "Hydrogenomics" (Grant No. JP18H05519).
Numerical calculations were performed using the facility of the Supercomputer Center, the Institute for Solid State Physics, the University of Tokyo

\nocite{*}

\end{document}


\title{Supplementary Material for \\
``Hydrogen-Induced Metal-Insulator Transition \\
Accompanied by Inter-Layer Charge Ordering in SmNiO$_3$''}
%

\author{Kunihiko Yamauchi}
\author{Ikutaro Hamada}%
\affiliation{%
Department of Precision Engineering, Graduate School of Engineering, Osaka University, Suita, Osaka 565-0871, Japan
}%

\date{\today}

\maketitle



\section{Examples of generated structure}

\begin{figure}[htb]
\begin{center}
\includegraphics[width=8.5cm]{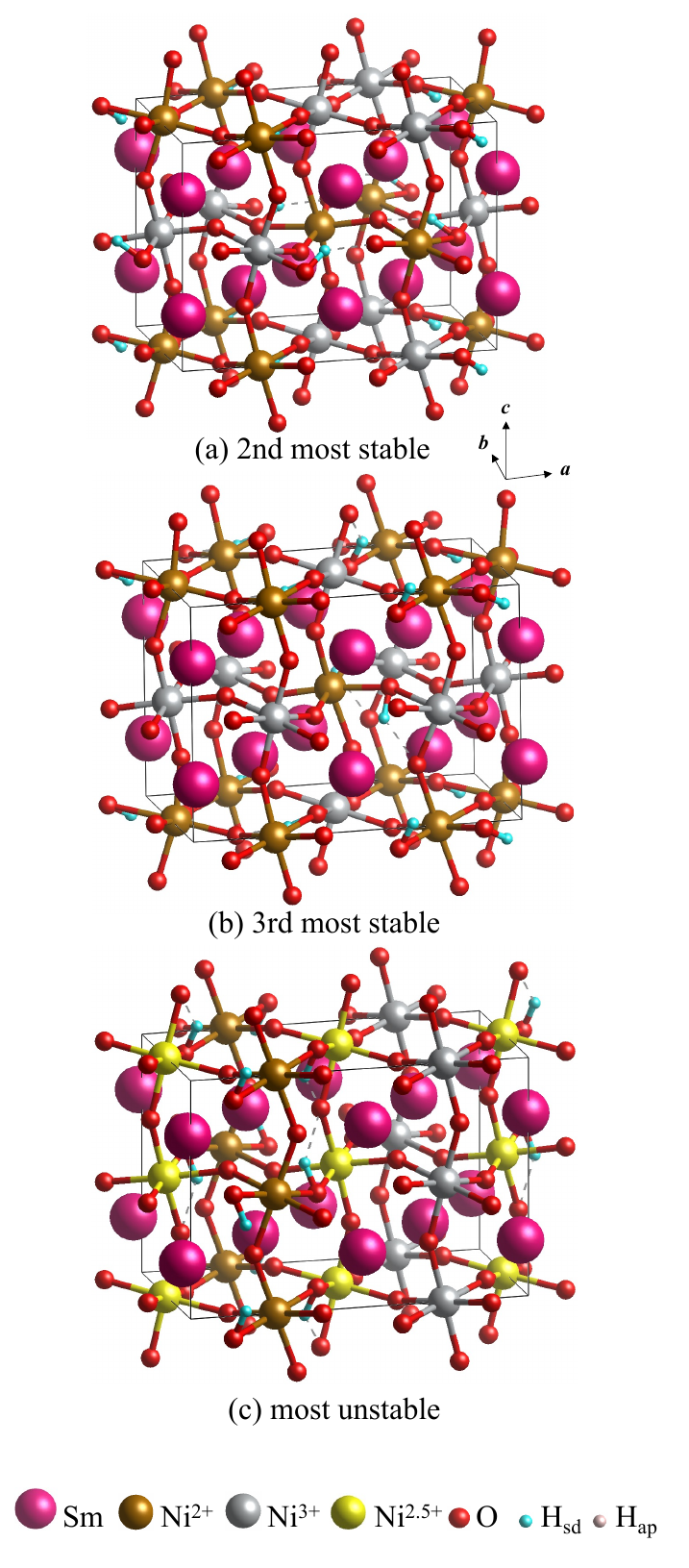}
\caption{\label{fig:crys_supl} 
(a, b) The second, the third most stable structure, and (c) the most unstable structure of 50\% H doped SmNiO$_3$. 
}
\end{center}
\end{figure}

Figure \ref{fig:crys_supl} shows some examples of the optimized atomic structure, {\it i.e.}, the second and the third most stable structures and the most {\it unstable} structure,  of 50\% H doped SmNiO$_3$. They show the formation energy as $-73.3$, $-72.4$, and $-44.7$ meV/H, respectively. Comparing them with the most stable 4H structure (see Fig. 1 (d) in the main text), it can be seen that the stable structures show the inhomogeneous H doping that leads to inter-layer-type Ni$^{2+}$/Ni$^{3+}$ charge ordering while the unstable structure shows rather homogeneous H doping that leads to charge disordered state with intermediate-valence Ni$^{2.5+}$ sites.

\section{Effect of lattice relaxation}

\begin{figure}[htb]
\begin{center}
\includegraphics[width=8.5cm]{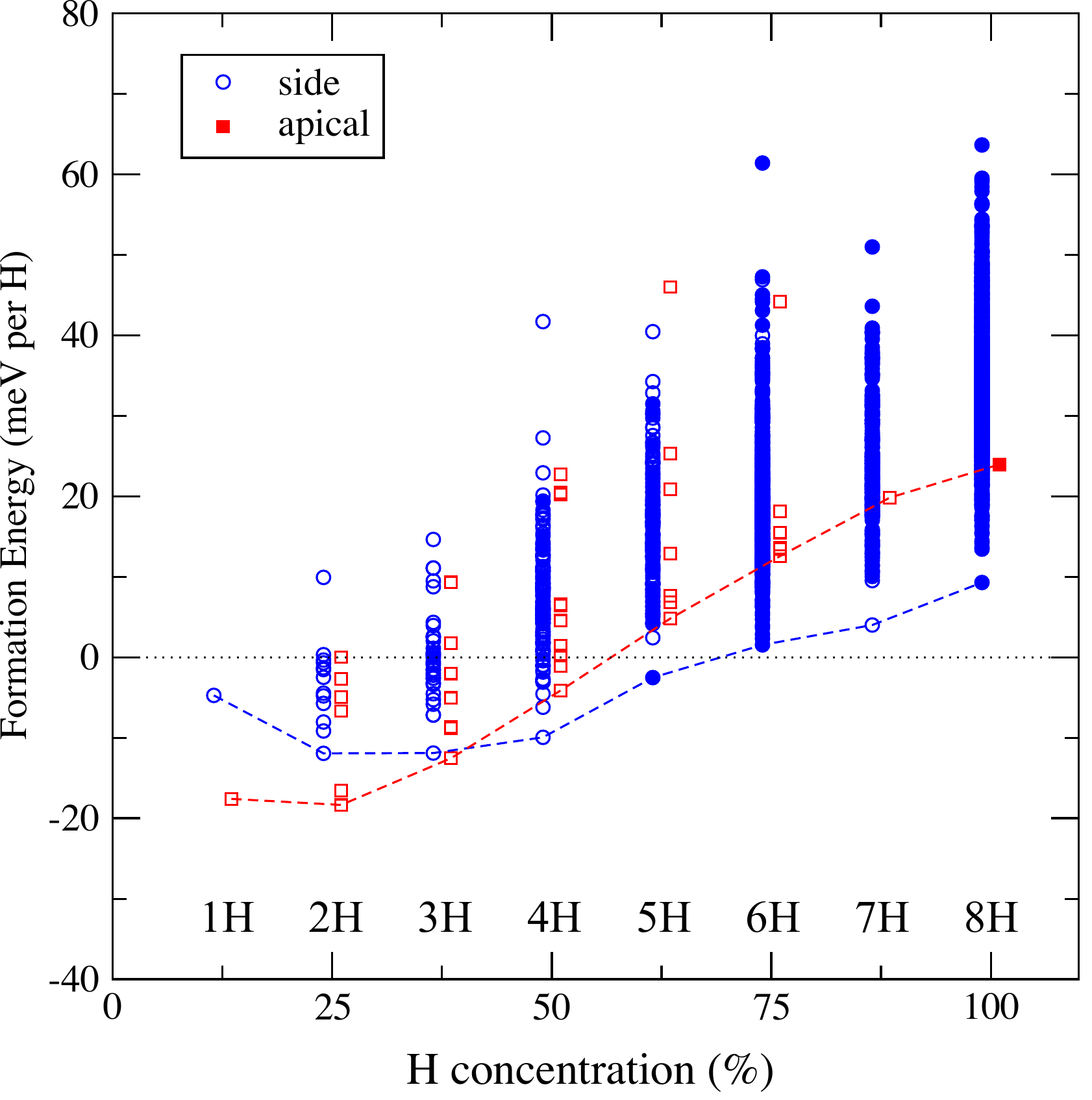}
\caption{\label{fig:en_strained} 
The formation energy of H$_x$-\sno\ as a function of the H doping concentration for side doping and apical doping. Open and closed symbols denote metallic and insulator states, respectively. In this calculation, the lattice parameters were fixed to those of non-doped SmNiO$_3$.
}
\end{center}
\end{figure} 

To investigate the effect of lattice relaxation, we repeated H-doping simulations in SmNiO$_3$ in which lattice parameters were fixed as those of a non-doped structure. As shown in Fig. \ref{fig:en_strained}, this resulted in a monotonous behavior of the formation energy with respect to the H concentration and the energy minima at 4H and 8H structures disappeared.

\section{Distortion modes}

\begin{figure*}[htb]
\begin{center}
\includegraphics[width=14cm]{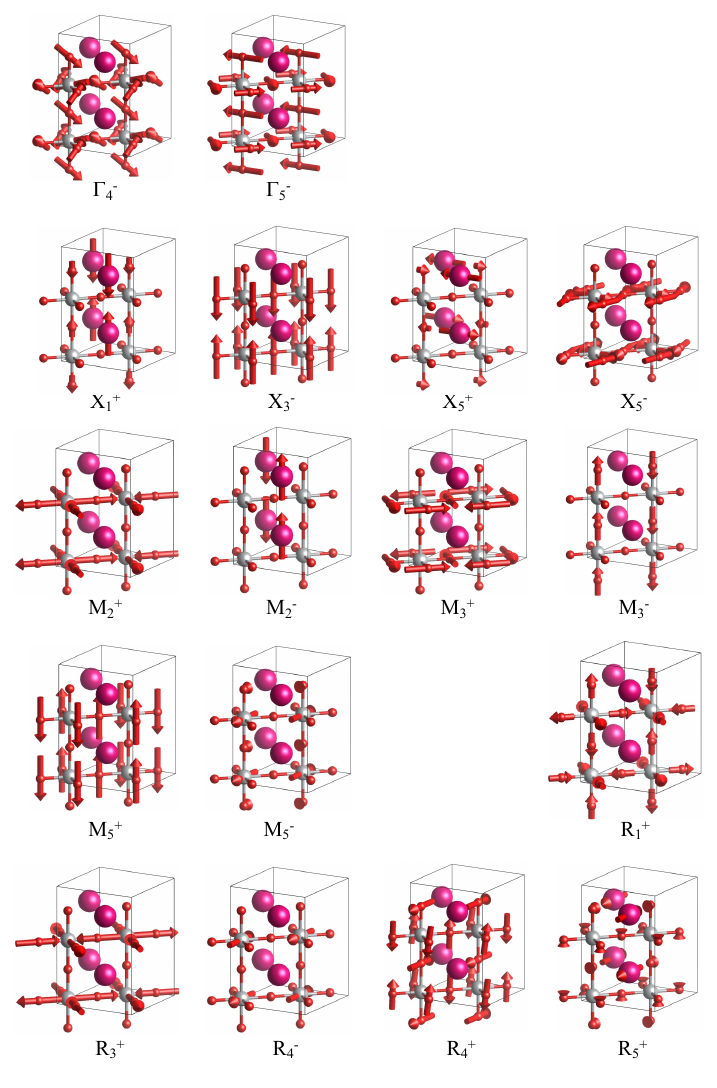}
\caption{\label{fig:distortion} 
Distortion modes for the $\sqrt{2}\times\sqrt{2}\times2$ \sno\ super cell with respect to the ideal cubic perovskite structure ($Pm\bar{3}m$). The mode names correspond to Tbl. I in the main text. 
}
\end{center}
\end{figure*}

Figure. \ref{fig:distortion} shows ionic distortion modes for \sno\ super cell. $X_1^{+}$ mode (interlayer breathing distortion) and $R_3^{+}$ mode (out-of-phase Jahn-Teller distortion) are responsible for the charge and orbital ordering in 4H structure. 

